\documentstyle[11pt,aasms4,flushrt]{article}

\lefthead{Nasiri \& Rezania}
\righthead{Quasars as Extreme Case of Galaxies}

\begin{document}

\title{Quasars as Extreme Case of Galaxies}

\author{S. Nasiri}
\affil{Center for Theoretical Physics and Mathematics, P. O. Box
11365-8486, Tehran, Iran\\
Department of Physics, Zanjan University,
Zanjan, Iran\\
Institute of Advanced Studies in Basic Sciences,
IASBS, Zanjan, P. O. Box 45195-159, Iran}

\author{V. Rezania}
\affil{Institute of Advanced Studies in Basic Sciences,
IASBS, Zanjan, P. O. Box 45195-159, Iran\\e-mail: rezania@sultan.iasbs.ac.ir}

\begin{abstract}
We introduce a phenomenological investigation of the evolution and large scale distribution of
quasars using a modified version of the Field and Colgate gravitational contraction model for
proto-galaxies. By studying the distribution of about 7000 quasars in 5 luminosity
classes, it seems that, such a model is capable of solving the energy problem
and discussing some of the observational properties of these objects.
A sketch of luminosity function of the quasars and the normal galaxies shows
a unified aspect for these objects.
The large scale
distribution of the quasars in the galactic coordinate shows the existence of filamentary
structures and voids in the same sence that have been resolved by exploring the
clusters of galaxies.
\end{abstract}


\section{Introduction}
In recent years, evidence has been mounting that quasars are extreme case of
galaxies rather than being truly different phenomena
(\cite{Ath96,Bat89,Pas89}).
One generally believes that the galaxies, as separate
units, originated
through some sort of gravitational instability. One assumes that, a fluctuation
in density either developed or pre-existed in the proto-galaxies  from which
the galaxies were to form. As a fluctuation grew in mass, it collapsed under the action
of gravity, cooled and eventually a galaxy was formed. If we assume that the quasars
are extreme case of galaxies we must seek for some characteristic
physical parameters which are responsible for the observational differences of these objects.
Field and Colgate (hereafter FC) (1976) considered the angular velocity of
proto-galaxies
as such characteristic parameter.  We will present a
simple approach to this model using the Tully-Fisher
relation (1977) and the Ogerell-Hossel phenonenological
formula (1991) to obtain a relation between  the luminosity and the angular velocity of the galaxies.
The FC model assumes that the size of the galaxies,
average mass of their constituent stars and their total energy output depend
on the rotation rate of the proto-galaxies or equally on the balance of the
gravity with the centrifugal force at the end of the contraction process. As an example, assume two proto-galaxies
with the same initial mass and size but, one with an
angular velocity ten times that of the other. The centrifugal force will then
be hundred times weaker for the slowly rotating proto-galaxy. Such an object
will eventually be about fifty times smaller in size and have, on the
average, stars of about a hundred times more massive than that of the fast
rotating one. Assuming that their
constituent stars are of main sequence type, the compact
object will generate about $2.5*10^3$ times more energy than the extended one.
According to the FC model,
the compact and the extended sources in the preceding example are the
representatives for a quasar and an ordinary galaxy, respectively.
Here we investigate the distribution of quasars in different luminosity classes together with the consideration
of their look back time. The result seems to agree with the FC
model along with, assuming the so called a "decay" mechanism for these objects.
Of course some people do not accept the FC model and consider
different scenario that assume the quasars to be the objects that
by some evolutionary mechanism become dimmer and dimmer in the course of time
(\cite{Kem97}) or, alternatively, to be a certain phase in the
process of the galaxies formation
(\cite{Hae93}). However, these propositions need, in turn, more
investigations and efforts.
Furthermore, we have considered the space distribution for about $40,000$
ordinary galaxies extracted from LEDA database (\cite{Pat96}). The
result looks
like the distribution of quasars associated with the lowest luminosity class.
In other word, it seems that the galaxies and quasars may be brought
under one and the same umbrella rather than being different phenomena.
The later claim is also supported by the behavior of the luminosity functions
of these objects. The luminosity function analysis is also employed to introduce
a critical angular velocity which specifies the branch point of the evolution
of proto-galaxies into the quasars and normal galaxies.
It is well known that the cluster of galaxies in space
are linked in a filamentary network
with the great voids between them (\cite{See94}). These observations concerning
the large scale structure of the universe may also be treated by studying
the distribution of the quasars. This is done and
a few filamentary structures and voids are resolved.
In section 2 the results and discussions
are given. Section 3 is devoted for concluding remarks.
\section{Results and Discussion}
If one assumes that the quasars are evolved from the slowly rotating proto-
galaxies and, therefore, possess much massive stars, one should accept that they must
evolve faster than the ordinary galaxies as well. Considering the previous example,
the quasars formed in this way would have a half life proportional to
the inverse square of its mass if presumably populated by the main sequence type of stars and will evolve about $10^4$
times faster than the corresponding ordinary galaxy.
Quasars including stars with the masses greater than eight solar mass, may eventually
disappear from the contact with the rest of the universe
as a result of collapsing after consuming their energy sources. This process, if done, may
lead to an evolutionary decay mechanism woking on these objects in the course of time.
To realize this phenomenon, one possible way is presumably to study the behavior
of the distribution of quasars in space. Considering the look back time of these
relatively distant objects, one expects a nonuniform distribution for them. In other words,
the plot of number density of quasars versus distance or in some sense,
versus time, should reveal more quasars at far distances (i. e. at very long
times ago) than in nearby regions. However, this is not satisfied by
Fig. 1, that shows more or less a uniform large scale density distribution for them.

We will return to this point later.
The required parameters for investigation of the quasars in the present work
are obtained using their absolute magnitudes
and redshifts given by Veron et al. (1991). Also the validity of Hublle's
law with the value 75$km/sec-Mpc$ for the Hublle constant is assumed. The completeness of data
is carried out by the well known $V/V_m$ method first used by Maarten Schmidt
to study the space distribution of a complete sample of radio quasars
from 3CR catalogue (\cite{Sch68}).
Let us come back to the FC model and introduce an alternative approach to it on the
basis of the Tully-Fisher relation which is
\begin{equation}
v=Al^{0.22},
\end
{equation}
where $l$ and $v$ are the luminosity and the circular rotation velocity of the galaxies,
respectively, and $A$ is a constant. On the other hand, one may consider the Oegerle-Hoessel phenomenological
formula as follows
\begin{equation}
r=Bv^{1.33}<SB>^{-o.83},
\end{equation}
where $r$ and $<SB>$ are the characteristic radius and mean surface brightness
of the galaxies, respectively, and $B$ is a constant. One then uses Eqs. (1) and (2) to derive
\begin{equation}
l=C\omega^{-0.7},
\end{equation}
where $\omega$ is the angular velocity and $C$ is a constant of proportionality.
Equation (3) implies that the luminosity increases as the angular velocity decreases
consistent with the imlication of the FC model. The corresponding data for known S0
morphological type of the galaxies is plotted in Fig. 2.

Here, the units of the luminosity
and the angular velocity are $10^{33}erg/sec$ and $10^{-15}rad/sec$, respectively.
Fitting a function of the form
$l=constant.\omega^{\alpha}$ to this figure gives the value $-0.78$ for $\alpha$,
satisfied by Eq. (3).
A question which arises here is, how the quasars with different angular velocities
evolve? According to FC model the quasars made of the proto-galaxies with lower
angular velocities will evolve faster than the normal galaxies made of the proto-galaxies with higher angular
velocities. Therefore, if one divides the full range of compelete sample of observed quasars into
different luminosity classes, one expects that the evolution behavior will look different for different classes
and the discrepancy encountered before, may be removed.
To do this we have classified the data
into 5 luminosity classes, in sucha way that, the luminosity increases with increasing the
order of the classes. The ranges of luminosity are not necessarily equal for each class and are chosen arbitrarily to be 0.0-0.5,
0.5-8.0, 8.0-40.5, 40.5-128.0 and 128.0-312.5 in the unit of $10^{45} erg/s$,
respectively. The corresponding distributions are plotted in Figs. 3
to 7. As an overal view, it is clear from these figures that the decay mechanism
is more pronounced for the luminous quasars as expected.
The space distribution of quasars associated with the 1st class, i. e. the dimmer
quasars, as plotted in Fig. 3, is almost uniform.

The situation demonstrates a group
of quasars that consume their energy sources by a relatively lower rate.
It means that the decaying process goes rather slowly for this class.
The situation is different for 2nd and 3rd classes as shown in Figs.
4 and 5. The slope of the distribution changes sign gradually for these classes,
while the quasars that are plotted in the left portion of the diagram
are gradually diminished. It may be interpreted in
such a way that the correspondig quasars produce energy with a higher rate
relative to those in the first class.

In Fig. 6, it is seen that the 4th luminosity class
of quasars have not been already observed at the distances less than about $2 Gpc$.
In other words, they have been disapeared before than about 6.5 billions of years ago
due to their relatively high rate of energy output. The situation is still more pronounced
for 5th class of the quasars that have been observed at the distances greater than
about $2.7 Gpc$ and, therefore, belonging to at least about 8.7 billions of years ago as shown in Fig. 7.
A rapid increase in the density of quasars at the right extremes in 4th
and 5th classes, seen in Figs. 6 and 7, may be considered to be responsible for a remarkable
increase in the density of observed quasars at large distance limit as shown in Fig. 1.
Therefore, one may conclude that the FC model modified by admitting the notion of decay
mechanism and the look back time is supported by observations.\\

By the same procedure we have investigated the distribution of a sample of about 40,000
normal galaxies in space. The result is plotted in Fig. 8. It looks like the distribution of quasars associated
with the first
luminosity class indicating probably a common origin for these objects.
As a supporting idea, the luminosity functions of the quasars and normal galaxies
are investigated and plotted in Fig. 9. It is seen from this figure
that, the absolute magnitude of observed quasars starts more or less from
a value, at which, that of the galaxies end up. This may be considered as another observational
evidence to consider these objects to be the same phenomena, but, with the different manifestations
starting from Fig. 3 and ending up at Fig. 8. Another observational result
which may be obtained from Fig. 9 is introducing a so called "critical angular
velocity". The proto-galaxies taking the angular velocities greater than this
critical value will eventually evolve to form the quasars and those with less than
this value will finally make the normal galaxies.
This is implied by Eq. (3), indicating a one to one correspondence between the
absolute magnitudes and the angular velocities. In addition, the peak values of the number of
quasars and galaxies may correspond to the so called "most probable angular velocities",
at which, the proto-galaxies tend to form the quasars and galaxies with the
highest rate.\\

As another result, one may consider the
distribution of the quasars, assuming them to be at the cosmological distances,
to recognize their corelation to form the large scale filamentary structures and voids.
They are intrinsically more luminous than the ordinary galaxies and could
be observed as farthest objects and, though, may be employed
to do a considerably deep exploration. In this respect
the entire sky map of the quasars are plotted in galactic coordinate as shown in Fig. 10.
The region of missing quasars is due to the dust clouds in our galaxy
which block our view of other quasars. Note the voids and clumpy
distribution of quasars looking like filaments. Their typical dimensions
are comparable to that of the "Great wall". As an example, the size of
one denoted by A with the approximate galactic lattitude and longitude of $+52^{\deg}$
and $-80^{\deg}$, respectively, is about 200 Mpc. Note some other structures
around the galactic north and south poles.\\

\section{Concluding Remarks}

A modified version of the FC model for governing the formation and evolution of quasars and galaxies
is considered.
The notion of the look back time and decay process for these objects
are considered as inherent properties of the model.
The logic behind it, however, is different. In particular, no
need arises to postulate the existance of the supermassive stars which is
not well understood yet. An arrangement of the quasars in
different luminosity classes and investigation of their evolution behveior
via each class provides a reasoable attempt to unify the origin of these objects with that of the
normal galaxies.
The unifying aspect is also recommended by investigating the luminosity function of these objects.
another intresting phenomenological idea is introducing the notion of critical
angular velocity. The protogalaxies may, eventually, evolve to quasars or galaxies depending
on whether their angular velocities are less or greater than the critical value.
Further, the most probable rate of formation of these objects seems to correspond
to the certain values of the angular velocities which may be obtained using Fig. 9
and Eq. (3).
A few filamentary structures and voids may be distinguished on the entire sky map
of the quasars plotted in the galactic coordinate. It seems that the quasars,
if to be at the cosmological distances, are suitable candidates to serve this purpose.\\

\acknowledgments

We are grateful to professor Y. Sobouti for his helpful comments.
We have made use of data from the Lyon-Meudon Extragalactic Database (LEDA)
compiled by the LEDA team at the CRAL-Observatoire de Lyon (France).Also we have made use of data from
Scientific Report of European Southern Observatory compiled by Veron, M. P. and Veron, P.\\

\clearpage

\figcaption {Variation of density of about 7000 quasars $(Gpc)^{-3}$ with distance $(Gpc)$.
 It shows almost a uniform distribution except for large distances,
 where a small increase occures. \label{Fig. 1}}
\figcaption {Variation of luminosity of the S0 type galaxies (in unit of $10^{33} erg/sec$)
 versus their angular velocities (in unit of $10^{-15} rad/sec$).  $f(x)$ is the best fit
function.  \label{Fig. 2}}
\figcaption {The same as Fig. 1 for the first luminosity class of quasars. It shows
 a more or less uniform distribution as expected for dimmer classes. \label{Fig. 3}}
\figcaption {The same as Fig. 1 for 2nd luminosity class of quasars. The slope of the
 distribution starts to increase indicating that the decay mechanism
 becomes important as the order of classes increases. \label{Fig. 4}}
\figcaption {The same as Fig. 1 for 3rd luminosity class of quasars. The density
 diminishes at the left (i. e. the nearby distances), while increases
 at the right (i. e. far distances) indicating the role of the decay
 of these objects due to their high rate output of energy. \label{Fig. 5}}
\figcaption {The same as Fig. 1 for 4th luminosity class of quasars. The decay mechanism
 is more pronounced here and the members of this class have not been already
 observed at the distances less than about $2 Gpc$. \label{Fig. 6}}
\figcaption {The same as Fig. 1 for 5th luminosity class of quasars. They are the most
 luminous quasars and, therefore, with the highest decay rate. They have not
 been already observed at the distances less than about 2.7$Gpc$. \label{Fig. 7}}
\figcaption {Variation of density of about 40000 ordinary galaxies $(Mpc)^{-3}$ with distance
$(Mpc)$.  The situation is simmilar to that
of the quasars associated with the first luminosity class indicating that
 the galaxies and quasars may be considered as different manifestations of
 the same phenomena. \label{Fig. 8}}
\figcaption {Distribution of quasars and galaxies in terms of their absolute magnitude.
 The broad function belongs to the quasars and the sharp one to the galaxies.\label{Fig. 9}}
\figcaption {Entire sky map of the quasars plotted in galactic coordinate showing some
 structural corelations. \label{Fig. 10}}

\setcounter{figure}{0}
\clearpage
\begin{figure}[h]
\vbox to2.6in{\rule{0pt}{2.6in}}
\includegraphics{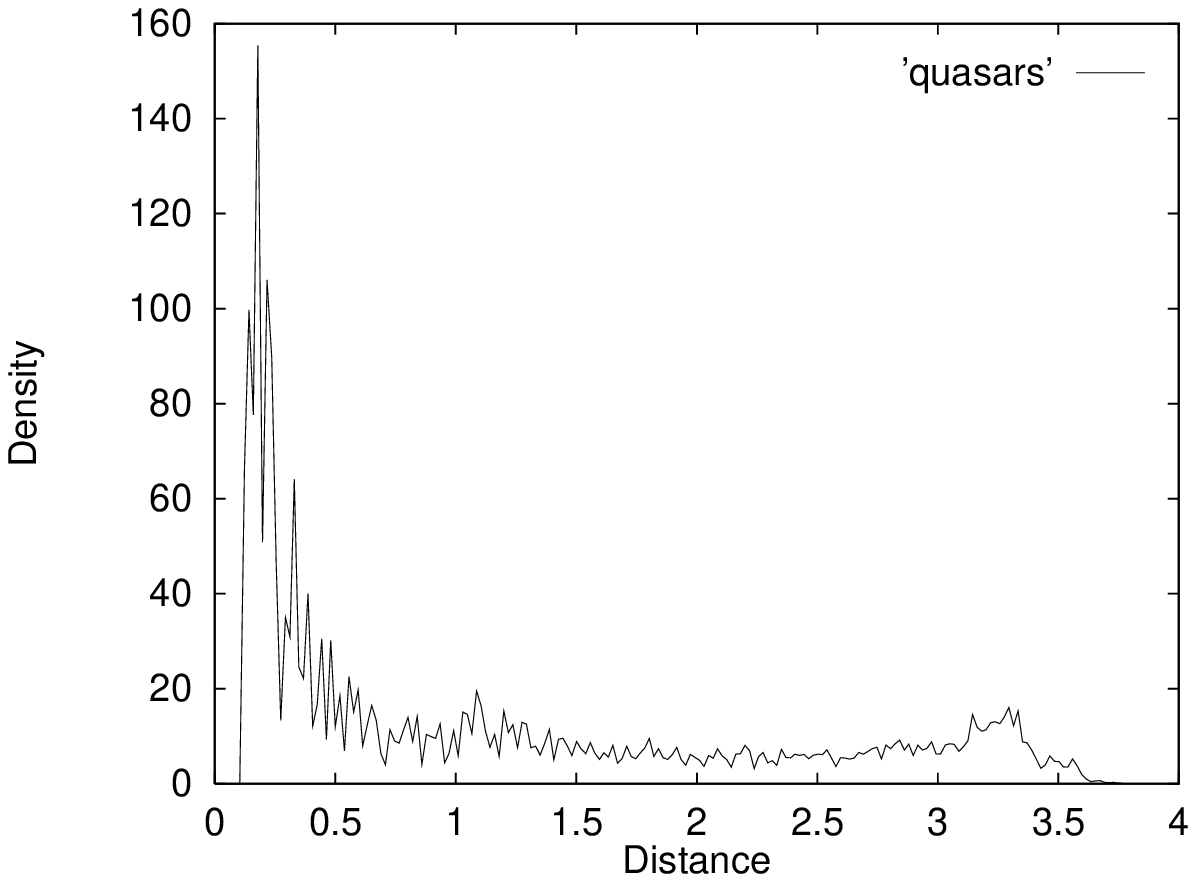}
\caption{}
\end{figure}

\clearpage
\begin{figure}[h]
\vbox to2.6in{\rule{0pt}{2.6in}}
\includegraphics{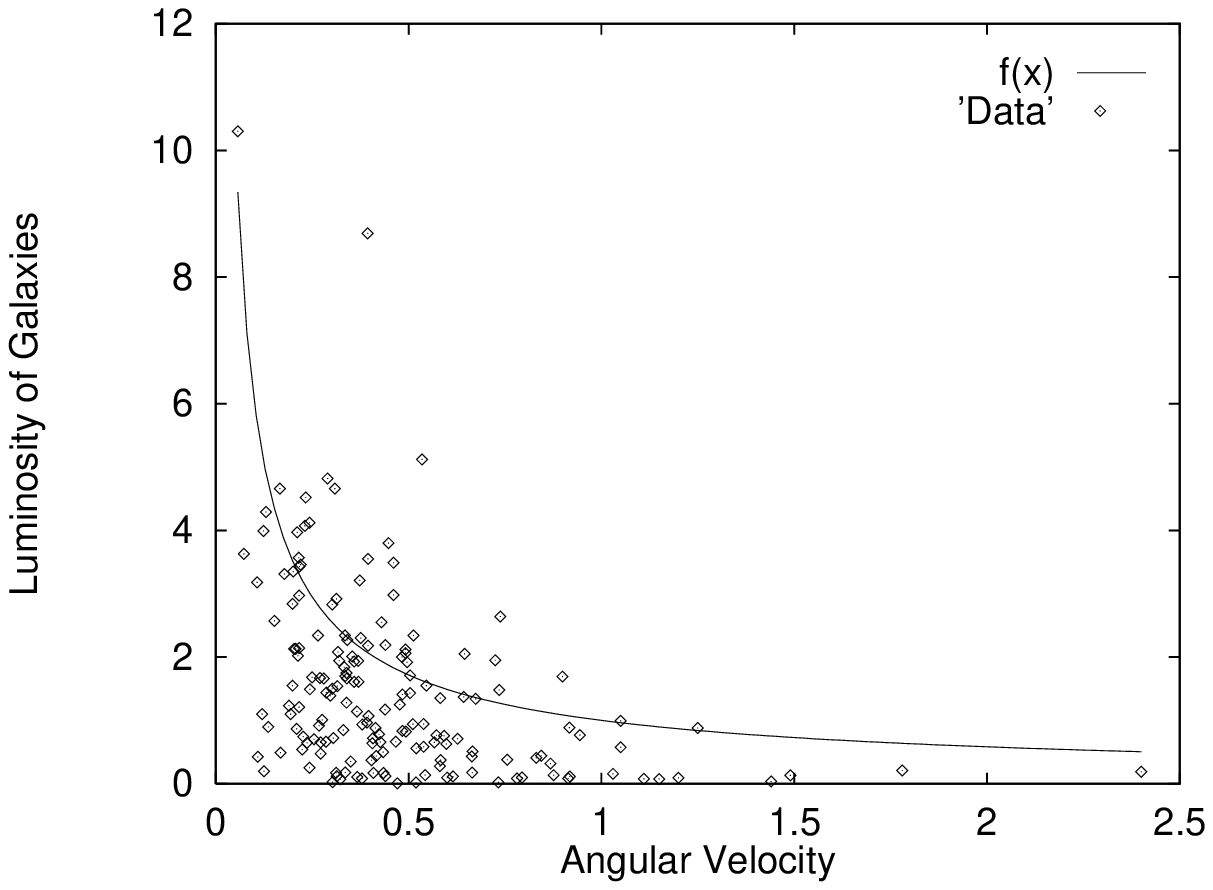}
\caption{}
\end{figure}
\clearpage
\begin{figure}[h]
\vbox to2.6in{\rule{0pt}{2.6in}}
\includegraphics{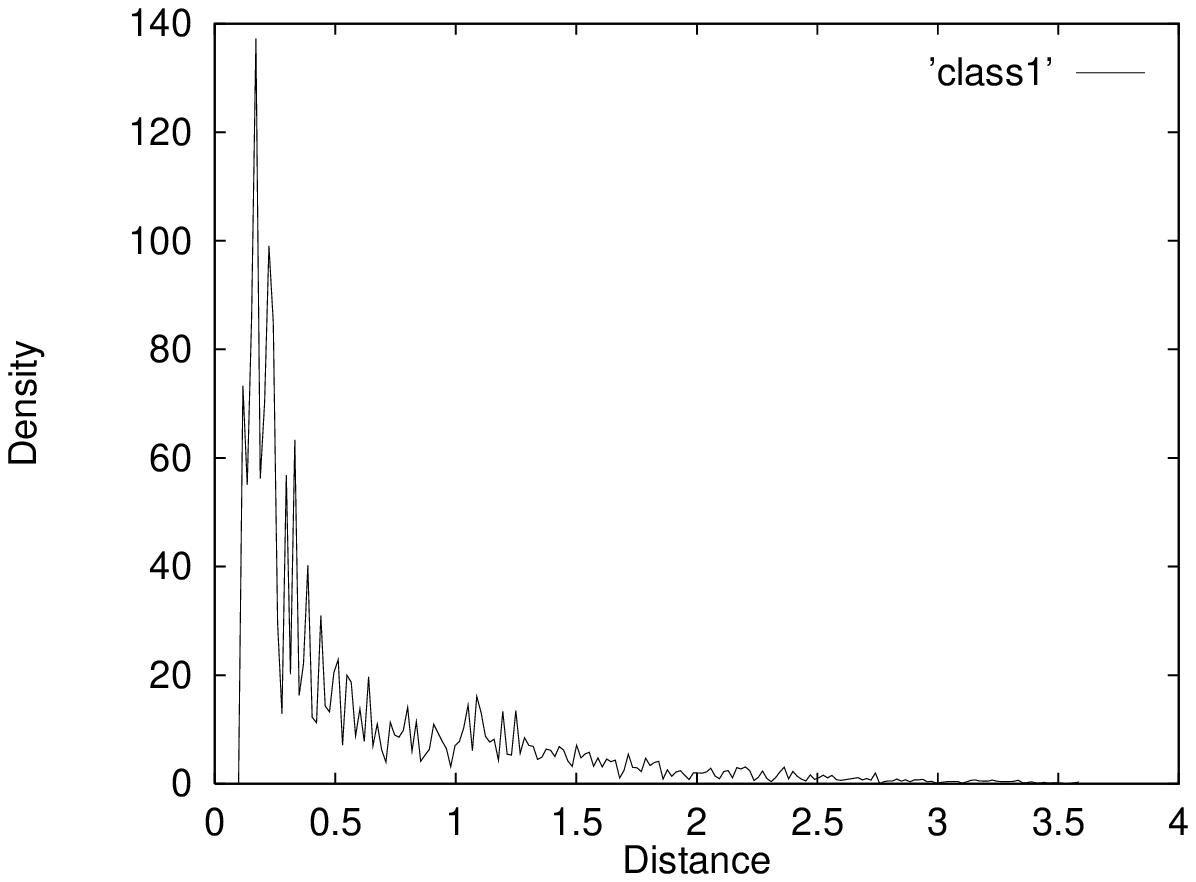}
\caption{}
\end{figure}
\clearpage
\begin{figure}[h]
\vbox to2.6in{\rule{0pt}{2.6in}}
\includegraphics{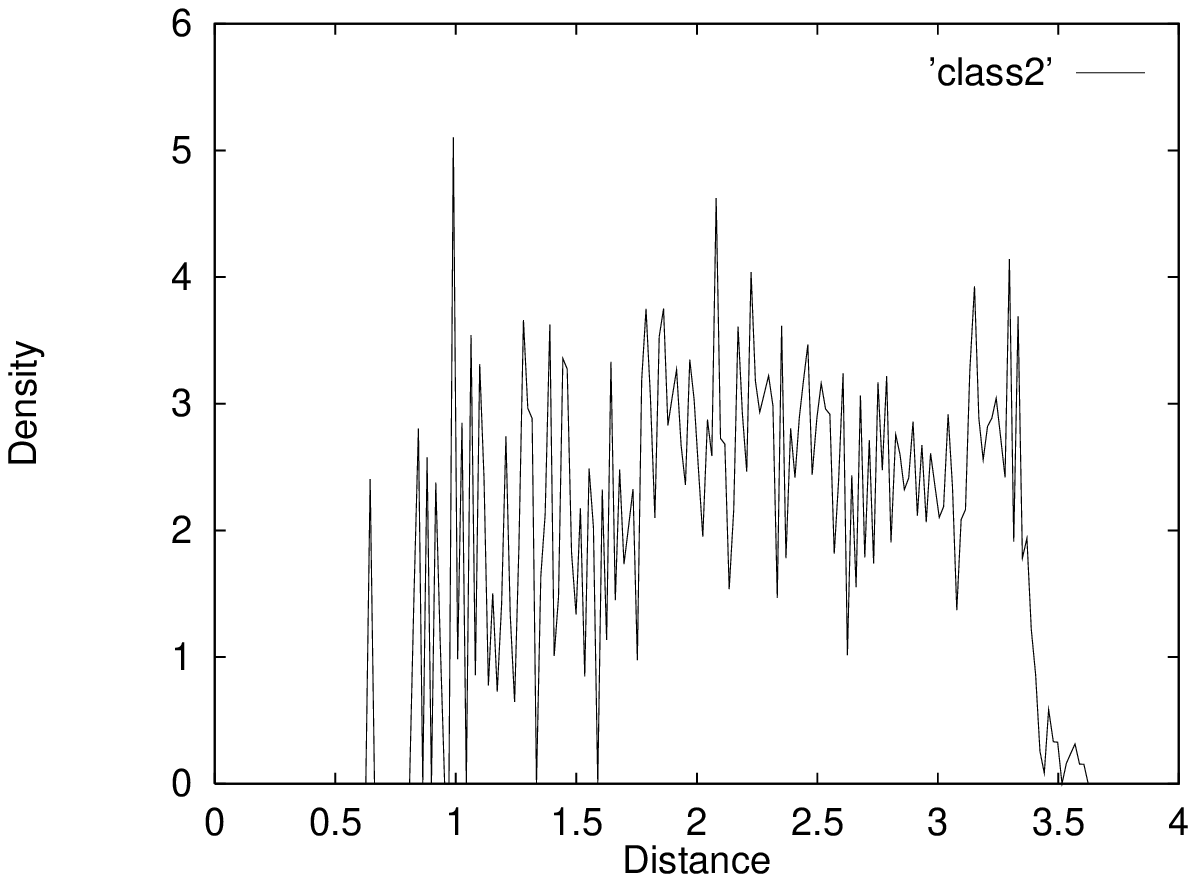}
\caption{}
\end{figure}
\clearpage
\begin{figure}[h]
\vbox to2.6in{\rule{0pt}{2.6in}}
\includegraphics{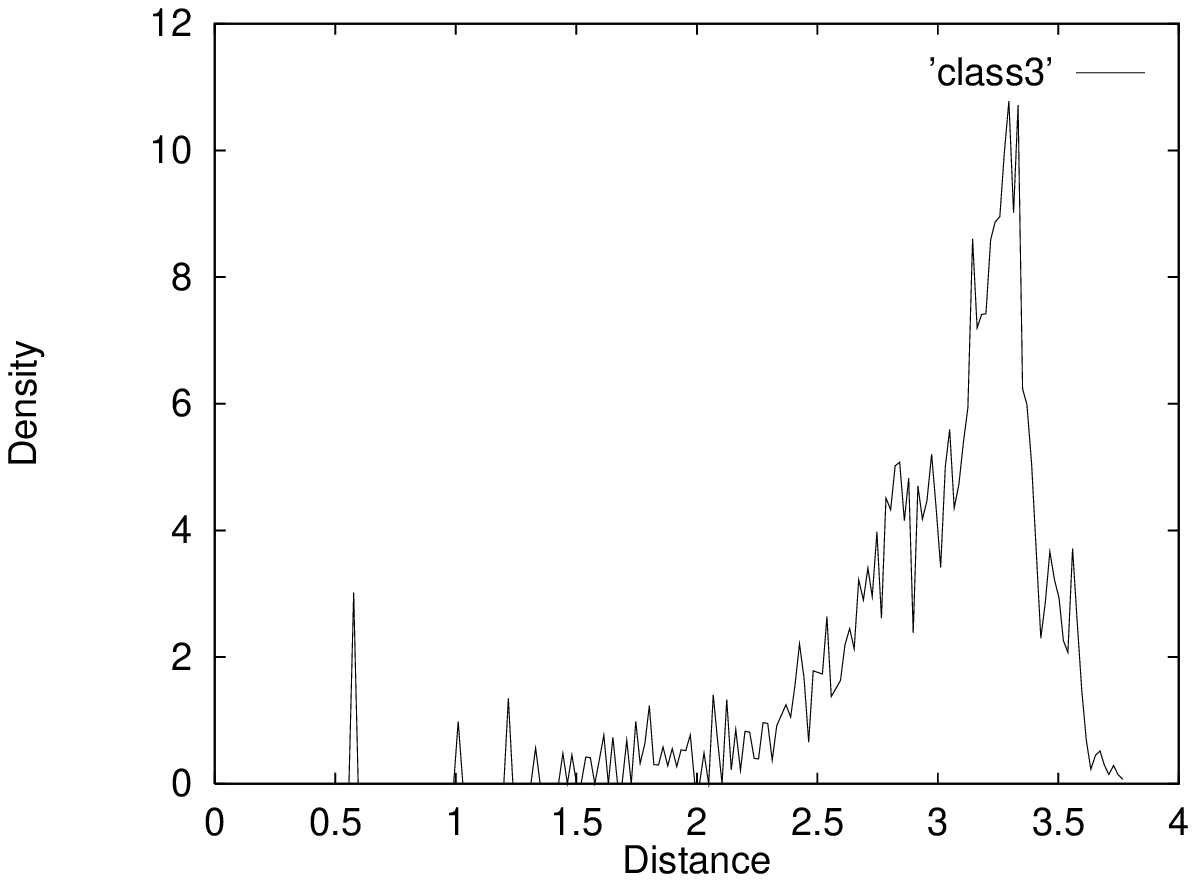}
\caption{}
\end{figure}
\clearpage
\begin{figure}[h]
\vbox to2.6in{\rule{0pt}{2.6in}}
\includegraphics{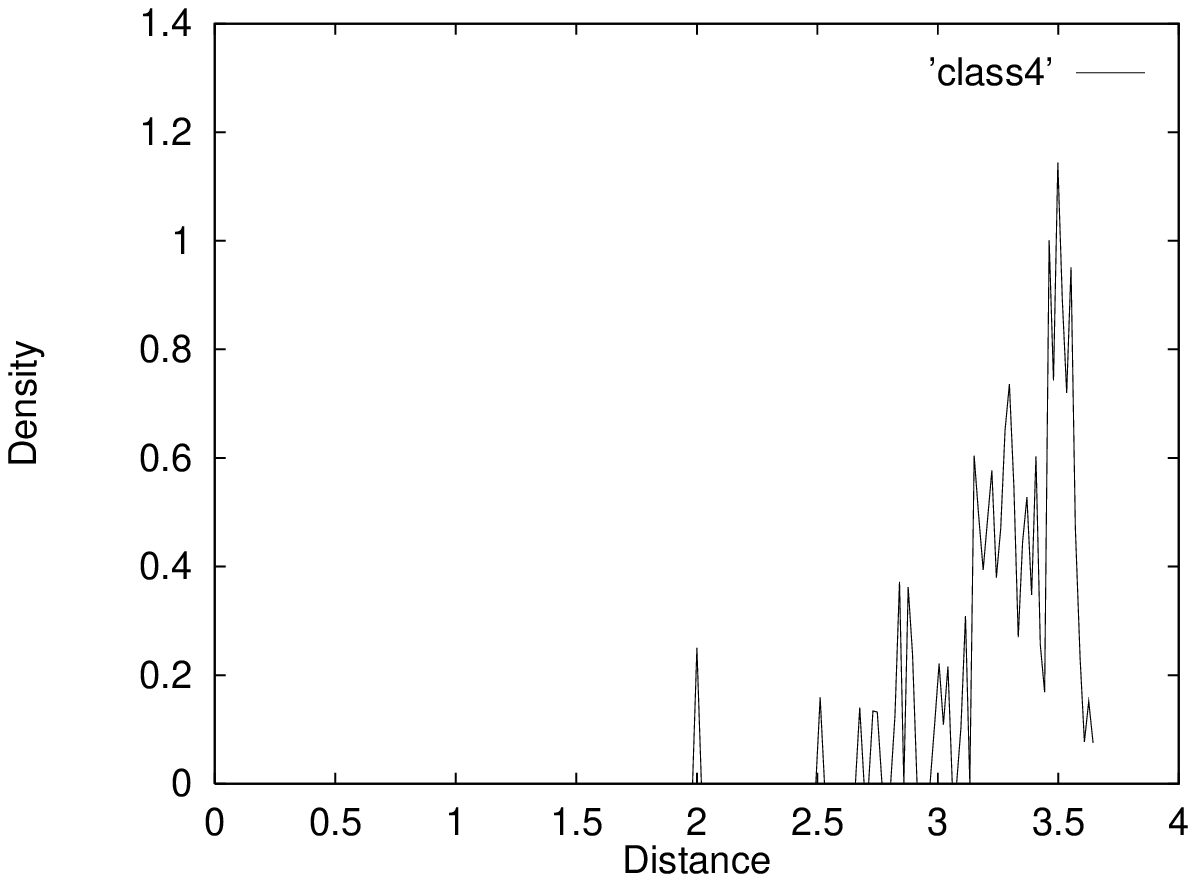}
\caption{}
\end{figure}
\clearpage
\begin{figure}[h]
\vbox to2.6in{\rule{0pt}{2.6in}}
\includegraphics{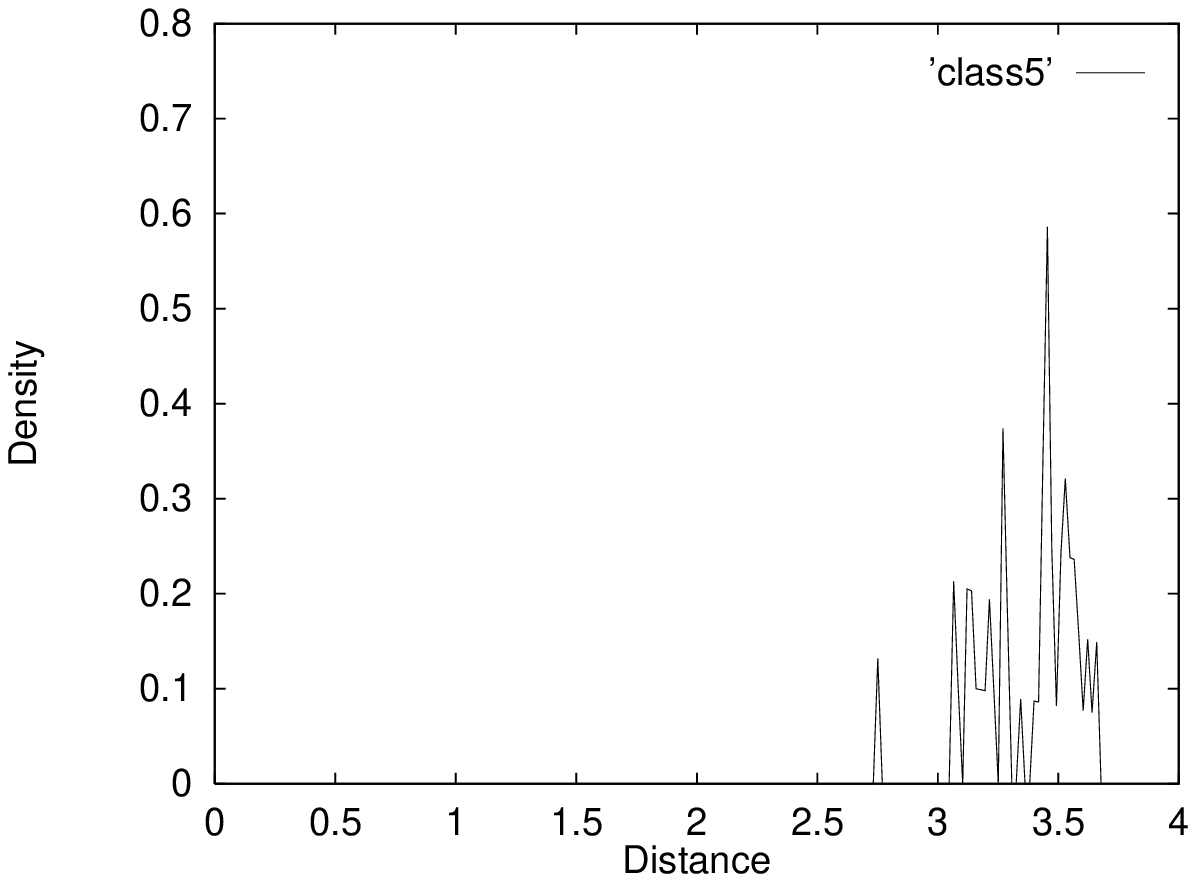}
\caption{}
\end{figure}

\clearpage
\begin{figure}[h]
\vbox to2.6in{\rule{0pt}{2.6in}}
\includegraphics{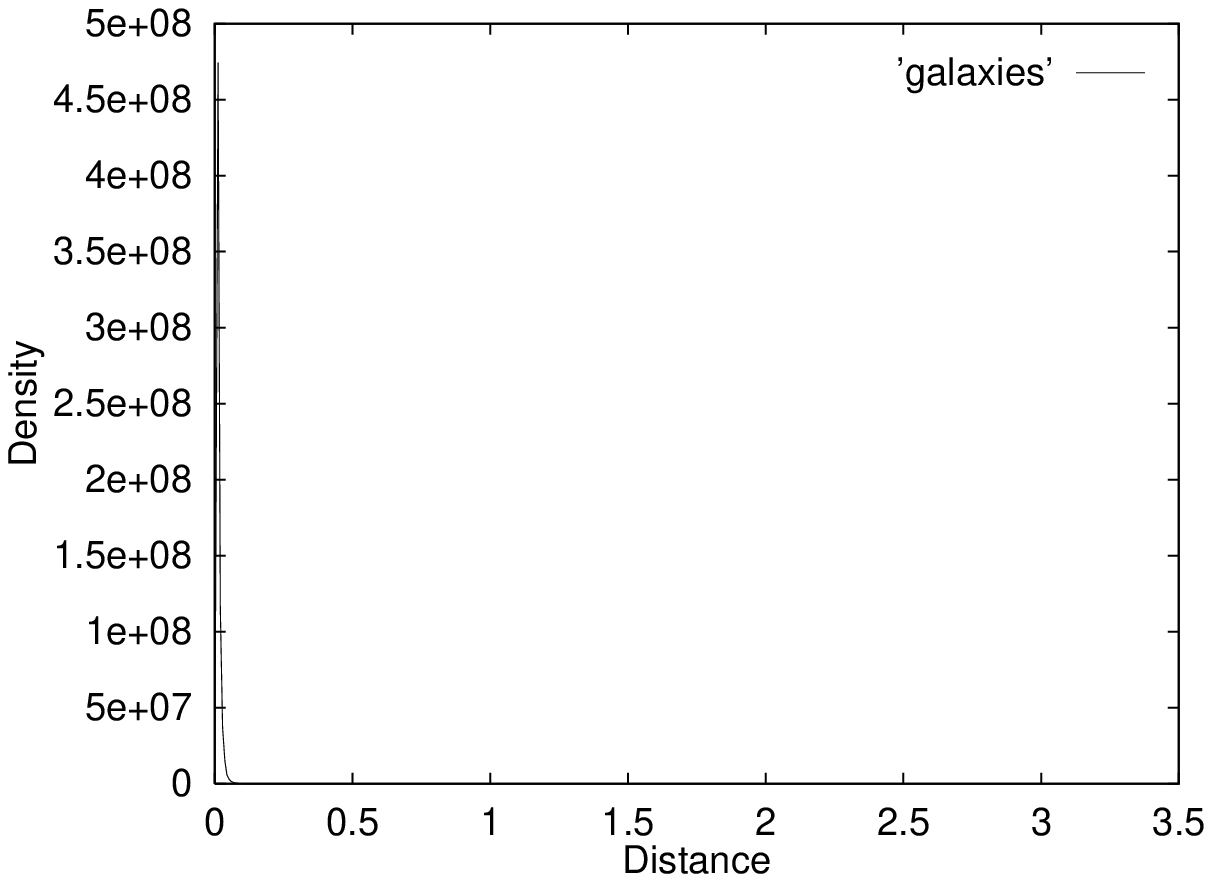}
\caption{}
\end{figure}
\clearpage
\begin{figure}[h]
\vbox to2.6in{\rule{0pt}{2.6in}}
\includegraphics{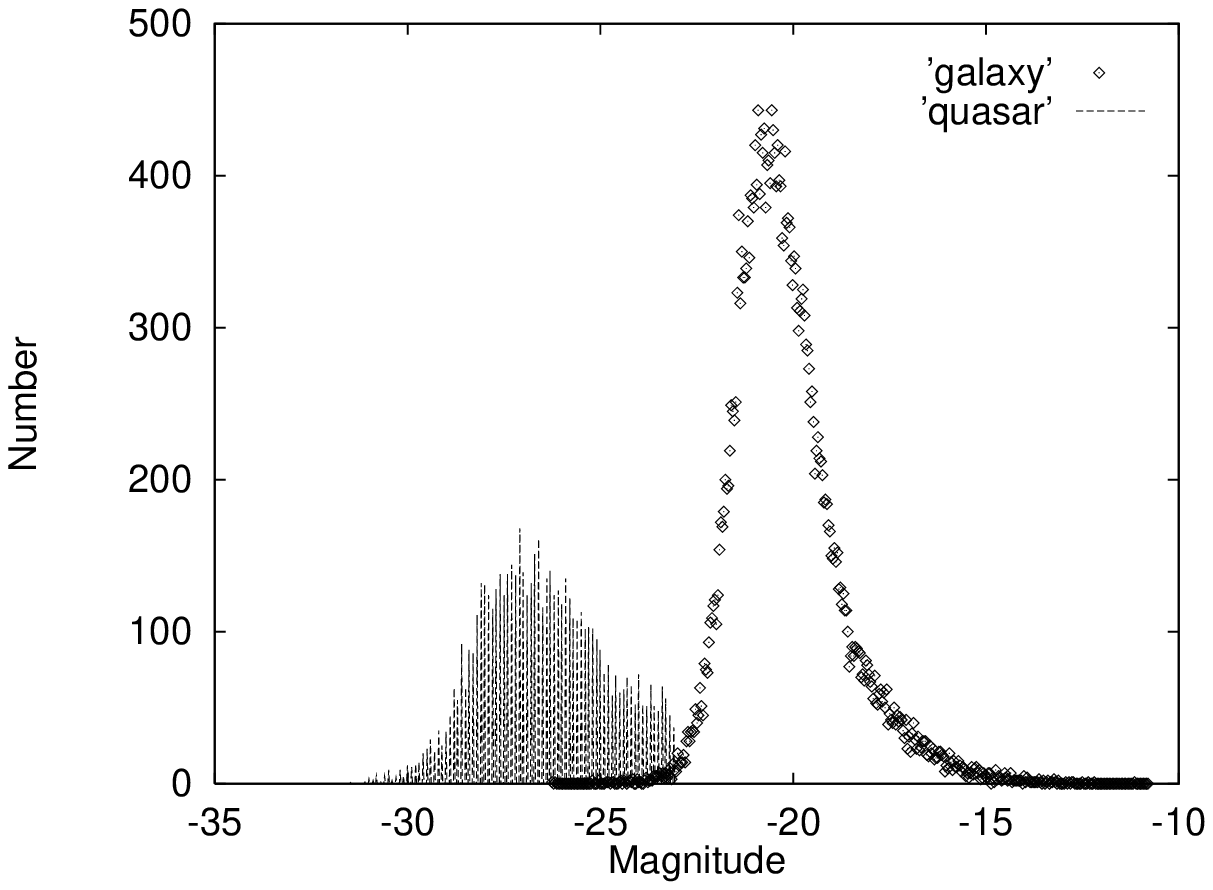}
\caption{}
\end{figure}

\clearpage
\begin{figure}[h]
\vbox to2.6in{\rule{0pt}{2.6in}}
\plotone{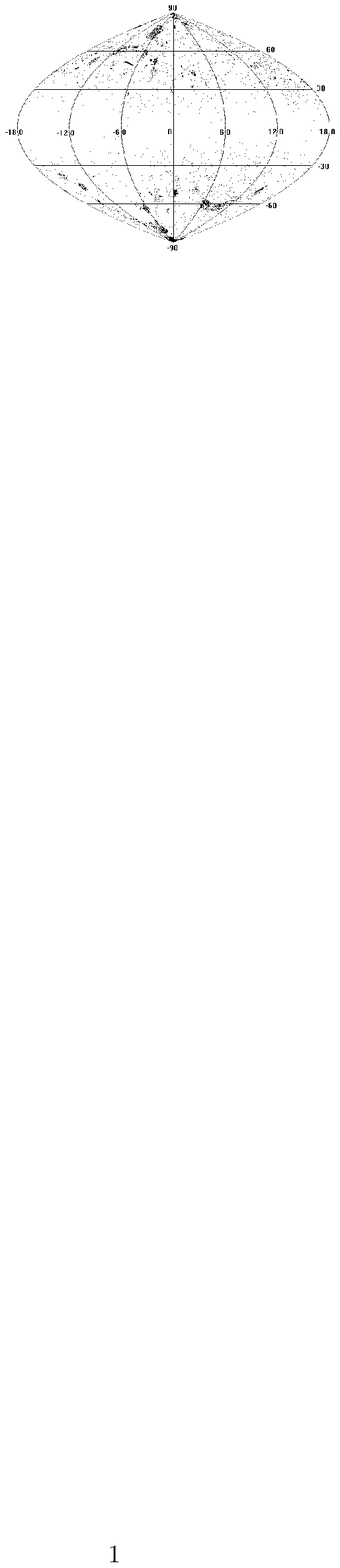}
\caption{}
\end{figure}

\end{document}